\documentstyle[12pt]{article}
\newlength{\extraspace}
\newlength{\extraspaces}
\newcommand{\bq}{\begin{eqnarray}
\addtolength{\abovedisplayskip}{\extraspaces}
\addtolength{\belowdisplayskip}{\extraspaces}
\addtolength{\abovedisplayshortskip}{\extraspace}
\addtolength{\belowdisplayshortskip}{\extraspace}}
\newcommand{\eq}{\end{eqnarray}}

\newcommand{\newsection}[1]
{\vspace{5mm}
\pagebreak[3]
\addtocounter{section}{1}
\setcounter{equation}{0}
\setcounter{subsection}{0}
\setcounter{footnote}{0}
\begin{flushleft}
{\large\bf \thesection. #1}
\end{flushleft}
\nopagebreak
\medskip
\nopagebreak}

\newcommand{\newsubsection}[1]{
 \vspace{5mm}
\pagebreak[3]
\addtocounter{subsection}{1}
\noindent{ \bf \thesubsection. #1}
\nopagebreak
\vspace{2mm}
\nopagebreak}

\newcommand{\ra}{\rightarrow}
\newcommand{\qa}{${\cal U}_{q}(sl(2))$~}
\newcommand{\qg}{$GL_{q} (2)$~}
\newcommand{\w}{$W_{\infty}$~}
\newcommand{\cL}{{\cal L}}
\newcommand{\va}{\vec{a}}
\newcommand{\vb}{\vec{b}}

\begin{document}

\addtolength{\baselineskip}{.8mm}

\thispagestyle{empty}

\begin{flushright}
{\sc PUPT}-1439\\
 January 1994\\
 hep-th/9401093
\end{flushright}
\vspace{.3cm}

\begin{center}
{\large\sc{Area-Preserving Diffeomorphisms,  $W_{\infty}$
 and ${\cal U}_{q}(sl(2))$ \\[3.5mm]
in Chern-Simons theory and Quantum Hall system} }\\[15mm].

{\sc  Ian I. Kogan}\footnote{ On leave of absence from
  ITEP, Moscow, 117259, Russia}\\[2mm]
{\it Physics Department, Jadwin  Hall\\[2mm]
      Princeton University,  Princeton, NJ 08544, USA} \\[15mm]

{\sc Abstract}
\end{center}

\noindent
 We discuss  a quantum  \qa symmetry in Landau problem,
  which  naturally arises due to the relation between
 the \qa  and the group of magnetic translations. The last one
  is connected with the \w and area-preserving (symplectic)
 diffeomorphisms which are the canonical transformations in the
  two-dimenssional phase space.
 We shall discuss the hidden quantum symmetry in
  a $2+1$ gauge theory with
 the Chern-Simons term and in a
  Quantum  Hall system which are both connected
 with the Landau problem.

\vfill
\newpage
\pagestyle{plain}
\setcounter{page}{1}
\stepcounter{subsection}
\newsection{Introduction.}
\renewcommand{\footnotesize}{\small}

Landau problem \cite{landau}, i.e. quantum mechanical description
of a charged particle with mass $m$ moving on the plane in a
 constant magnetic field $B$ normal to the plane, has important
 applications  in
 different areas of theoretical physics.  The spectrum of one-body
 problem consists of degenerate  energy levels - the so called
 Landau level. The degeneracy of each level proportional to
 total magnetic  flux  and  this system  exibits a lot
 of interesting properties.

This problem is the cornerstone  of the
 quantum Hall effect (QHE) description
   \cite{qhe}.
 Landau problem also arises \cite{km} in
 a topologically massive gauge theory \cite{tmgt},  i.e.
 $2+1$-dimensional gauge theory  with the Chern-Simons term,
 quantization.
It  was shown \cite{km}, \cite{cslandau}  that the  Hilbert space
 of the theory is a direct  product of the massive gauge particles
 Hilbert space (one free massive particle in the most simple
 $U(1)$ case)
 and  some quantum-mechanical Hilbert space. In the
  $U(1)$ case this quantum-mechanical  Hilbert space  is the product of the
  $g$ copies (for a genus $g$ Riemann surface) of the
 Hilbert space for the  Landau problem on the torus.
  In the infinite mass limit
  all levels except the first one are  decoupled  as well
 as massive particles Hilbert space and we have only
 the first Landau level which becomes the Hilbert space
 of the topological  Chern-Simons theory \cite{witten}.

 Landau Hamiltonian is not invariant under the translation
 group, however
  it was known for a long time that it is invariant
   under  the group of magnetic
 translations  \cite{magtransl}.
  Recently  Wiegmann and Zabrodin \cite{wz}  demonstrated that
 magnetic translations can be expressed through  generators
 of the quantum algebra \qa  \cite{sklyanin}, \cite{qg}
  and applied this representation
  to formulate the Bethe-Ansatz for the
 Azbel-Hofstadter problem, i.e. the problem of Bloch electron
  in magnetic field. The Bethe-Ansatz solution was generalised
 in \cite{fk}.  Let us note that several years ago
  Floratos \cite{floratos}
    constructed the representations  of the quantum group
  \qg with $q = \exp(2\pi i/N)$  using the $N\times N$   representations
   of the Heisenberg group algebra, which were equivalent to
 the representations of the magnetic translations.

    It was also discussed recently that there is a \w symmetry \cite{w}
 in the Landau problem \cite{kogan} - \cite{ctz}. This
 symmetry was discussed  in a context of a
  topologically massive gauge theory (TMGT) in \cite{kogan} and
 of a  quantum Hall effect (QHE) in \cite{sakita}, \cite{ctz}.
 It is amusing that the \w algebras are connected with
  an  algebra of area-preserving
 (or symplectic)  diffeomorphisms of the two-dimensional manifolds,
  for example it is an  infinite-dimensional algebra    of canonical
 transformation  in a two-dimensional phase space $(p,q)$.  It is also
 a  symmetry of the relativistic
 membranes after gauge fixing - this symmetry and it connection with
 the $SU(\infty)$ were considered in \cite{membrane}, \cite{ffz}.

The aim of this paper is to discuss the connection between
  quantum symmetry, magnetic translations  and  area-preserving
   diffeomorphisms in Landau problem and to discuss the
  \qa symmetry in $2+1$ gauge theories with Chern-Simons terms
 and quantum  Hall systems. Let us note that  for general $q$
 the irreducible representations of  \qa   are qualitatively the
 same as in $sl(2)$ case \cite{neroot}. However in the case
 $q^{n} = 1$ there is  only finite number of
irreducible representations \cite{root} and the new classification
 of the states in TMGT and quantum  Hall systems exists.

 The organization of the paper is as follows. In the  next  section,
  which bears essentially a review character, we
 consider the Landau problem and discuss the relations
 between  canonical transformations, magnetic translations
 and the quantum algebra \qa. We shall demonstrate how after
 the restriction on the first Landau level the symmetry
 under the area-preserving
   diffeomorphisms will appear. We shall demonstrate how the
 this symmetry is connected with quantum group symmetry and
 will discuss  the construction of quantum algebra \qa
 from the group of magnetic translations. At the end of this
 section  this construction
  will be  generalaized to the case of
 Fairlie-Fletcher-Zachos triginometric algebras
 with nontrivial central extension \cite{ffz}, which can not be reduced
 to  any magnetic translations.   In  section $3$  the
 canonical quantization and Landau levels picture in
 topologically massive gauge theory will be considered and
 the action of the quantum  group on the Hilbert space will
 be obtained.   The corresponding  problem, as  we shall show, is
 equivalent to the Landau problem on a torus and we shall consider
 several examples of  different representations of \qa.
 It will be shown that the natural value for
 deformation parameter $q = \exp( 4\pi i/k)$ where $k$ is the
 Chern-Simons coefficient
    In the section  4 we shall consider  the
 quantum group symmetry in a  quantum  Hall system. We shall
   remember how \w (to be more precise $W_{1+\infty}$
 appears in a quantum Hall system due to
 the incompressibility of the ground state and then
  generalize our construction of \w and
  \qa on many-body case and will present arguments in favour
 of deformation parameter $q = \exp(2\pi i \nu)$, where
 $\nu$ is a filling factor. We shall also consider the
 action of \qa on ground state and lowest excitations.
 In conclusion the  obtain results will be discussed  as well
 as some  possibilities to  find the quantum group symmetry
 in other physical models invariant under the area-preserving
 diffeomorphisms and \w, such as  $c=1$ strings and
 two-dimensional  Yang-Mills theory.

\newsection{\w, magnetic translations and \qa in Landau problem}
\newsubsection{Landau problem}

The action for  Landau problem, i.e. for  particle with mass $m$ moving
 on the $(x_{1},x_{2})$ plane in a magnetic field $B$ is
\bq
S_{L} = \frac{m}{2} \int  \dot{x}_{i}^{2}dt + \int  A_{i}\dot{x}_{i} dt =
\frac{m}{2} \int  \dot{x}_{i}^{2} dt
 + \frac{B}{2}\int (x_{2}dx_{1}-x_{1}dx_{2})
\label{Laction}
\eq
where  we choose a
  vector potential in a symmetric gauge  $A_{i} = (B/2)\epsilon_{ij}x_{j}$.

    Let us consider the configuration and phase spaces of this
 problem. The
 canonical momenta:
\bq
p_{i} = \frac{\partial L}{\partial \dot{x}_{i}} =
 m \dot{x}_{i}  + \frac{B}{2}\epsilon_{ij}x_{j}
\label{Lmomentum}
\eq
with the usual commutation relations (or Poisson brackets
 in the classical limit $[,] \rightarrow -i\{,\}$)
\bq
[p_{i},p_{j}] =  [x_{i}, x_{j}] = 0;
\;\;\;[p_{i}, x_{j}] = -i\delta_{ij}
\eq
The canonical Landau Hamiltonian and the eigenvalues $E_{n}$ are
\bq
H = \frac{1}{2m}\left(p_{i} - \frac{B}{2}\epsilon_{ij}x_{j}\right)^{2}
 = -\frac{1}{2m}\left(\frac{\partial}{\partial x_{i}} -
i\frac{B}{2} \epsilon_{ij}x_{j}\right)^{2} \nonumber \\
E_{n} = (n +
1/2) \frac{B}{m}~~~~~~~~~~~~~~~~~~~~~~~~~~~~~~~~~~
\label{Lhamiltonian}
\eq

Let us
 notice  that  Hamiltonian  (\ref{Lhamiltonian})  depends
 effectively only on  two coordinates in the phase space, not four.
 Introducing the variables $a$ and $a^{+}$ depending on $(p_{i}- A_{i})$ only
\bq  a^{+} = \left(p_{1} - \frac{B}{2}x_{2}\right) +
i  \left(p_{2} + \frac{B}{2}x_{1}\right) = 2p_{\bar{z}} +
 i \frac{B}{2} z \nonumber \\
a =   \left(p_{1} - \frac{B}{2}x_{2}\right) -
i  \left(p_{2} + \frac{B}{2}x_{1}\right) =
2p_{z} -
 i \frac{B}{2} \bar{z}
\label{La}
\eq
 one gets the Hamiltonian
\bq
 H = \frac{1}{4m} (aa^{+} + a^{+}a), ~~~~~~~~[a,a^{+}] =  2B
\label{LH}
\eq
  Here  $z(\bar{z}) = x_{1}\pm i x_{2}$ and
$p_{z,\; \bar{z}}= -i\partial/\partial z(\bar{z})$
 are the corresponding  conjugate momenta.  There is
 another pair  $b$ and $b^{+}$, which depends on $(p_{i}+A_{i})$,
 commuting with $a$ and $a^{+}$
\bq
 b^{+} = \left(p_{1} +\frac{B}{2}x_{2}\right) -
i  \left(p_{2} - \frac{B}{2}x_{1}\right) =
2p_{z} +
 i \frac{B}{2}\bar{z} \nonumber \\
b =  \left(p_{1} +\frac{B}{2}x_{2}\right) +
i  \left(p_{2} - \frac{B}{2}x_{1}\right) =
 2p_{\bar{z}} -
 i \frac{B}{2} z
\label{Ldefinitionb}
\eq
with commutation relation $ [b,b^{+}] =  2B$.
 The angular momentum operator can be written as
\bq
J = b^{+}b - a^{+}a
\eq
One can see that  $b^{+}$ and $a$ increase and $b$ and $a^{+}$ decrease
 the angular momentum.

It is easy to see that the states on the first Landau level $|1>$   are
 annihilated by $a$ and has the form
\bq
a|1> = 0, ~~~~~~ <z,\bar{z}|1> = \Psi_{1}(z,\bar{z}) =
 \Phi(\bar{z}) \exp(-\frac{B}{4}z\bar{z})
\label{1level}
\eq
 where $\Phi(\bar{z})$ is an arbitrary antiholomorphic function
 and operators $b$ and $b^{+}$ do not change the level number when acting
 on states at a given level.  For a state with angular momentum $l$
 one has $ \Phi_{l}(\bar{z}) = \bar{z}^{l}$. One can consider another
 basis, parametrized by a momentum in $x_{1}$ (or any other) direction
\bq
\Psi_{1}(p|\vec{x}) =
\exp\left(-i\frac{B}{2}x_{1}x_{2} + iBp x_{1}\right)
\exp\left(-\frac{B}{2}(x_{2} - p)^{2}\right) = \nonumber \\
\exp(-\frac{B}{4}z\bar{z})
\Phi_{p}(\bar{z}) = \exp(-\frac{B}{4}z\bar{z})
 \exp\left(-\frac{B}{2}p^{2} + iBp \bar{z} + \frac{B}{4}\bar{z}^{2}\right)
\label{momentumwf}
\eq
where momentum is defined as $Bp$.

 Let us note that one can consider the  restriction on the first Landau level
  taking the limit
 $m \rightarrow  0$. In this limit one gets from
  (\ref{Lmomentum})
\bq
2p_{\bar{z}} = p_{1} + i p_{2} =
  -i\frac{B}{2}z;~~2p_{z} = p_{1} - i p_{2} =
  i\frac{B}{2}\bar{z}
 \eq
and then
\bq
 a = a^{+} = 0; ~~~~~~ b^{+} =i B\bar{z};~~~~~~
 b =  -
 i B z
\label{aabb}
\eq
 The physical  meaning of  this reduction is the follows -
 operators  $a$ and $a^{+}$
 acting on the state at the given level $n$ shift it to $n\mp 1$.
 To be at a first  level  we  must put these operators
 to zero after which   $b$ and $b^{+}$
 play the role of the coordinate on the reduced  phase space.
 Let us also note that the action (\ref{Laction}) in the
 limit $m \ra 0$ transforms into
 \bq
S_{m=0} = \frac{B}{2} \int  (x_{1}\dot{x}_{2} - x_{2}\dot{x}_{1}) dt =
  \frac{B}{2}\int (x_{2}dx_{1}-x_{1}dx_{2})
\label{Lactionzerom}
\eq
and one can easliy see that $x_{1}$ and $x_{2}$ becomes  the coordinates on
 the phase space, the action  in the case of the closed
 trajectories proportional to the area
 $A=\oint (x_{2}dx_{1}-x_{1}dx_{2})$ and is invariant under the
 action of the area-preserving diffeomorphisms. The last one
 are nothing but a canonical transformation on the first Landau level.
 Let us also note that the connection between Chern-Simons theory and
 Landau
 problem in the limit $m \rightarrow 0$, i.e. reduction on the
 lowest level,  where discussed in  \cite{dj}.

\newsubsection{Canonical transformation on the first Landau level,
 \w  algebra and the group of magnetic translations}

 Let us remember
 some well-known facts about canonical transformations (see,
 for example \cite{arnland}). By
 definition canonical transformations are
   diffeomorphisms
 of the phase space which preserve the symplectic structure
 $\omega = \sum_{p} dq^{l}\wedge dp^{l}$.

The canonical transformations are usually  defined by the
 generation function depending   on both  old
 ( $p$ or $q$) and new ($P$ and $Q$) phase space
 coordinates, for example one can consider arbitrary
 $F(q,Q)$ and put
 \bq  p_{i} = \partial F/\partial q_{i};~~~
 P_{i} = -\partial F/\partial Q_{i}
\eq
 It is easy to see that $P,Q$ are new canonical coordinates.
There is however another representation, namely one can consider
 evolution with respect to some "Hamiltonian" $W(p,q)$
(which is an arbitrary
 function on the phase space and has  nothing common with the
 physical Hamiltonian). The change
 in quantities $p$ and $q$ during this evolution may itself be
 regarded as a series of canonical transformations.  Let
 $p$ and $q$ be the values of the canonical variables at time $t$
 and $P$ and $Q$ are their values at another time $t + \tau$.
 The latter are some function of the former, depending on $\tau$
  as on parameter
\bq
Q = Q(q,p;\tau),~~~~~~P=P(q,p;\tau)
\eq
These formulae can be considered as the canonical transformation
from the old coordinates $p,q$ to the new ones $P,Q$. This
 representation is convenient for the infinitesimal transformation,
 when $\tau \rightarrow 0$. In this case using Hamiltonian
 equations of motion with "Hamiltonian" $W(p,q)/\tau$ one gets
\bq
 Q_{i} = q_{i} + \dot{q}_{i}\tau = q_{i} + \{q_{i}, W\}; ~~~~~
 P_{i} = p_{i} + \dot{p}_{i}\tau = p_{i} + \{p_{i}, W\}
\label{can1}
\eq
 where
\bq
\{A, B \} = \sum_{i} \frac{\partial A}{\partial q_{i}}
\frac{\partial B}{\partial p_{i}} -\frac{\partial B}{\partial q_{i}}
\frac{\partial A}{\partial p_{i}}
\label{brackets1}
 \eq
 is the usual Poisson brackets.

 The canonical
 transformations acting on the two-dimensional phase
 space $(q,p)$  are defined by
\bq
 \delta q =  \{q, W(p,q)\}=\frac{\partial W(p,q)}{\partial p}
 ; ~~~~~
 \delta p   =  \{p , W(p,q)\}= -\frac{\partial W(p,q)}{\partial q}
\label{can}
\eq
 where $W(p,q)$ is an arbitrary function. The fact that these
 transformations preserves the area can be easily   checked
 using    the fact that the general  infinitesimal area-preserving
 diffeomorphism takes the form
\bq
x_{i} \rightarrow x_{i} + \xi_{i}(x);~~~\partial_{i}\xi_{i} = 0
 \eq
 where $x_{i} = (q,p)$.
 General solution  of   $\partial_{i}\xi_{i} = 0$   is the sum of the
 two terms
\bq
\xi_{i}(x) = \epsilon_{ij}\partial_{j}W(x) + \sum_{a=1}^{b_{1}}
c_{a}{\bf \xi}^{a}_{i}
\eq
 where  first term  describes infinite number (all possible functions
 $W(\xi)$)  of the local
 co-exact solutions and the second term describes the finite number
(given by the first Betti number $b_{1}$) of the harmonic forms
 on two-dimensional phase space. It is easy to see that diffeomorphisms
 generated by the first term are nothing
but canonical transformations (\ref{can}).

  Any function $f$ on the phase space is transformed under the
 canonical transformation generated by  $W$ according to the rule
 $ w f =\delta_{W} f = \{ f, W \}$, where $w$ is the operator
 corresponding to function $W(\xi)$.
  Using the Jacobi identity
 $\{\{f,W_{1}\} W_{2}\} - \{\{f, W_{2}\} W_{1}\} + \{\{W_{1},W_{2}\}f\}=0$
 one can check that
  algebra of the  area-preserving diffeomorphisms is given
 by the Poisson brackets
\bq
 [w_{1}, w_{2}] f  =
  [\delta_{W_{1}},\delta_{W_{2}}]f = \delta_{\{ W_{1},W_{2}\} } f
\eq
 Any function $W$ can be written in terms of the complete set of harmonics
\bq
W_{\vec{n}} = exp (i \vec{n}\vec{x})
\label{W}
\eq
where $\vec{n} = (n_{1}, n_{2})$ with real $n_{1},n_{2}$ in the case
 when the phase space is a plane. If the phase space is a torus
 (as it will be in the case of TMGT or QHE on torus) one has
 integers $n_{1},n_{2}$
  if one defines a  torus phase space $T^{2}$ to be a square with both
 sides equal to $2\pi$.
 One gets the commutation relations for operators $w_{\vec{n}}$
 computing the  Poisson bracket for $W_{\vec{n}}$, \cite{membrane}
  \bq
[w_{\vec{m}}, w_{\vec{n}}] = (\vec{m}\times\vec{n})~ w_{\vec{m} + \vec{n}}
\label{w1}
\eq
where $\vec{a}\times\vec{b} = a_{1}b_{2} - a_{2}b_{1}$.
 One can see that  (\ref{w1}) is
 nothing but the commutation
 relations for the  classical $w_{\infty} $ algebra \cite{w}.

 Let us now consider canonical transformations  acting
 on the Landau problem phase space. The general canonical
 transformations are acting on the whole four-dimensional
 phase space  and after quantization they will mix different
 Landau levels.  However there is a  special subgroup of
 the canonical transformations acting  on the two-dimensional
 subspace of the phase space generated by  commuting with the Hamiltonian.
 This means that this transformations do not mix different
 Landau levels and thus acting on each Landau level as on
 two-dimensional phase space. It is evident that generators
 of this area-preserving (symplectic) transformations will
 depend on $b$ and $b^{+}$ (see(\ref{Ldefinitionb})) which
 commute with  Hamiltonian (\ref{LH}).

 After quantization we get instead of (\ref{W}) the quantum version
 \bq
{\cal W}_{n,\bar{n}} =
\exp\left(\frac{1}{2}(n b^{+} - \bar{n} b)\right) ~
  =
\exp(\frac{1}{2} n b^{+}) \exp(-\frac{1}{2} \bar{n} b)
\exp(-\frac{B}{4}n\bar{n});
\nonumber \\
 \bigl[ {\cal W}_{n,\bar{n}},{\cal W}_{m,\bar{m}}  \bigr] =
 -2i\sin\frac{B}{2}(n_1m_2 - n_2 m_1)
 {\cal W}_{n+m,\overline{n+m}} ~~~~~~~~~~~~~~
\label{W1}
\eq
 Here $n(\bar{n}) = n_{1}\pm  i n_{2}$  and
 the classical limit corresponds to  weak magnetic field
 $B \rightarrow 0$  after  obvious rescaling of ${\cal W}_{n,\bar{n}}
 \ra B w_{n,\bar{n}}$.
 For integer $n$ and $m$ the algebra (\ref{W1})
  is  the Fairlie-Fletcher-Zachos (FFZ) trigonometric
 algebra
  \cite{ffz}

Let us note that this is precisely the algebra of magnetic translations
\cite{magtransl} which  (in a gauge $A_{i} = (B/2) \epsilon_{ij} x_{j}$)
  is generated by the operators\footnote{ Let us note that the
 definition of magnetic translations is gauge dependent -
 $\vec{\nabla} +i \vec{A}$ commute with a Hamiltonian only in a
 symmetric  $A_{i} = (B/2) \epsilon_{ij} x_{j}$ gauge.
 In general case one must  add some gauge dependent terms
  in (\ref{mtranslation}). One can show,
 however, that
 the algebra of  magnetic translation does not depend on gauge and later
 we shall work only in the  symmetric gauge.}
\bq
T_{\vec{\xi}} = \exp\left(\vec{\xi}
\left(\vec{\nabla} +i \vec{A}\right)\right),
{}~~~~~~~~~~~
T_{\vec{\xi}}T_{\vec{\eta}} = \exp\left(-i\frac{B}{2}(\vec{\xi}\times
\vec{\eta})\right)~T_{\vec{\xi}+\vec{\eta}}
.\label{mtranslation}
\eq
Substituting (\ref{Ldefinitionb}) into (\ref{W1}) one  find that
\bq
{\cal W}_{n,\bar{n}} = T_{\vec{\xi}}, ~~~~~~\xi_{i} = \epsilon_{ij}n_{j}
\eq
and the action on the first level wave functions (\ref{1level}) is
 as  follows:
\bq
T_{\vec{\xi}}~ \Psi_{1}(\vec{x}) =
\exp\left(i\frac{B}{2}(\vec{\xi} \times \vec{x})\right)
 \Psi_{1}(\vec{x} + \vec{\xi}) =
 \exp\left(\frac{B}{4}(\bar{\xi} z - \xi \bar{z})\right)
 \Psi_{1}(z+\xi,\bar{z}+\bar{\xi})
\label{Taction}
\eq
where $\xi = \xi_{1} + i\xi_{2} = -in, \bar{\xi} =
 \xi_{1} - i\xi_{2} = i\bar{n}$ and vector notation is
 obvious.  This action  has especially simple form in a case
 of wave functions (\ref{momentumwf})
\bq
T_{\vec{\xi}}~ \Psi_{1}(p|\vec{x})
 = \exp\left(- i \frac{B}{2}{\xi}_{1}{\xi}_{2}
 + iBp {\xi}_{1} \right) \Psi_{1}(p-{\xi}_{2}|\vec{x})
\label{pTaction}
\eq
Later we shall use  a notation $|p>$ for wave function
$ \Psi_{1}(p|\vec{x})$.

The FFZ  algebra (\ref{W1})  commutes  with the
  Hamiltonian  $H  \sim (a a^{+} + a^{+} a)$ and thus
 acts independently on each Landau level.
  One can construct another
 \w algebra from $a,a^{+}$  operators  which we shall call
 $\tilde{W}_{\infty}$ and
  gets: \w $ \otimes \tilde{W}_{\infty}$.
 The first algebra acts on each Landau level, the second
 one (tilde) algebra mixes the level and acts in a simple form
onto the coherent states
 $|\alpha>_{r} \sim \exp{(\alpha a^{+})}|0>_{r}$.

 One can consider  another form of generators  \cite{sakita}, \cite{ctz}
 \bq
\cL_{n,m} = (b^{+})^{n+1}b^{m+1},~~~~~n,m \geq -1
\label{Lnm}
\eq
 with commutation relations
\bq
[\cL_{n,m}, \cL_{k,l}] = 2B\left((m+1)(k+1) - (n+1)(l+1)\right)\cL_{n+k,m+l}
 + O(B^{2})
\label{comLnm}
\eq

 This algebra is called $W_{1+\infty}$  in the literature
 (see \cite{ctz} and references therein).
  After obvious rescaling one has classical
 $w_{\infty}$ in the limit $B \ra 0$.
 It is easy to see that  expanding  generators
 ${\cal W}_{n,\bar{n}}$  (\ref{W1})  in $n, \bar{n}$ one gets
  generators $\cL_{n,m}$  as  expansion coefficients.
\bq
{\cal W}_{n,\bar{n}} =  \exp(-\frac{B}{4}n\bar{n})
 \sum_{k,l = 0}^{\infty} (-)^{l} \frac{n^{k}}{2^{k} k!}
\frac{\bar{n}^{l}}{2^{l} l!} \cL_{k-1,l-1}
\label{WLconnection}
\eq

Let us note that FFZ algebra (\ref{W1})   is a Weyl-Moyal \cite{wm}
 deformation of the Poisson-Lie algebra with usual Poisson brackets
$\{f,g\} = \partial_{z}f \partial_{\bar{z}}g
- \partial_{\bar{z}}f  \partial_{z}g$. The Moyal bracket $\{,\}_{M}$
 is defined as follows:
\bq
\{f,g\}_{M} = \sum_{s=0}^{\infty} \frac{(-)^{s}B^{2s}}{(2s+1)!}
 \sum_{j=0}^{2s+1}(-j)
  \left(\begin{array}{c} 2s+1 \\ j \end{array}\right)
 \left[ \partial_{z}^{j} \partial_{\bar{z}}^{2s+1-j} f\right]
 \left[ \partial_{z}^{2s+1 - j} \partial_{\bar{z}}^{j} g\right]
\label{moyal}
\eq
It is easy to check for $W_{n,\bar{n}} = \exp[1/2(nz-\bar{n}\bar{z})]$
 one gets the classical $w_{\infty}$ algebra (\ref{w1})
 for usual Poisson brackets $\{,\}$
  and  FFZ algebra (\ref{W1}) for Moyal brackets $\{,\}_{M}$.

\newsubsection{\qg and \qa in Landau problem}

 There is a natural connection between the  FFZ algebra  (\ref{w1})
 \cite{manin}, \cite{mitya} and quantum group \qg. Let us consider
 the quantum plane
\bq
 UV = q VU
\label{qplane}
\eq
 and introduce a quantum group  $GL_{q}(2)$ which is  defined as $2\times 2$
 matrices
\bq
 L = \left( \begin{array}{cc}
a & b \\
 & \\
 c & d \\ \end{array}\right)
\eq
where the matrix elements $a,b,c,d$ obey relations
\bq
ab = q^{-1} ba, ~~~~~ ac = q^{-1} ca,~~~~~~~  cd = q^{-1} dc, \nonumber \\
bd = q^{-1} db, ~~~~~ bc = cb, ~~~~~~~~~~~ ad - da = (q^{-1} - q) bc
\eq
It was shown in \cite{manin} that \qg is the group of automorphisms of the
 quantum plane (\ref{qplane}), i.e. $U'V' = q V'U'$, where $U'$ and $V'$
  are the images of the action of \qg on $U$ and $V$.  It is easy to see
  that
 $T_{\vec{n}} = q^{n_{1}n_{2}}U^{n_{1}}V^{n_{2}}$  with $q = \exp(iB)$
 are generators of the FFZ algebra and  thus  \qg  naturally acts
 on this algebra preserving  the commutation relations (\ref{W1}).

  It is amusing that one can construct  also quantum algebra \qa  from
 the elements of the magnetic translations group (\ref{mtranslation})
 (see \cite{wz},\cite{fk}  and references therein).
 The commutation relations of \qa are defined as follows \cite{sklyanin},
 \cite{qg}:
\bq
q^{J_{3}} J_{\pm} q^{-J_{3}} = q^{\pm 1} J_{\pm}  \nonumber \\
\left[ J_{+}, J_{-}\right] = \frac{q^{2J_{3}} - q^{-2J_{3}}}{q - q^{-1}}
\label{qa}
\eq
where the first relation equivalent indeed to usual commutation relation
 $[J_{3},J_{\pm}] = \pm J_{\pm}$. This algebra is transformed into
 an ordinary Lie algebra $sl(2)$ in the limit $q \ra 1$.

 The construction of the $J_{\pm}$ and $J_{3}$ generators from
 the given magnetic translations group (\ref{mtranslation}) is not
 unique. One can  get arbitrary value for a deformation
 parameter $q$ in a general case. However  later we shall
 demonsrate that in physically interesting situations like
 quantum Hall effect or $2+1$ dimensional gauge theory the
 choice of $J_{\pm}$ and $J_{3}$ will be dictated by additional
  physical arguments  and there   will be some "natural"
  choice of parameters.

  Let us    present      the following  construction
 depending on  two  arbitrary
 noncollinear vectors  $\va$ and $\vb$ on a plane  and
 four complex parametres $\alpha, \beta, \gamma, \delta$.
 Considering the
 following superpositions of magnetic translation generators:
\bq
J_{+} = \frac{1}{q - q^{-1}} \left(\alpha~T_{\va} +
\beta~ T_{\vb}\right), ~~~~
J_{-} =  \frac{1}{q - q^{-1}} \left(\gamma~T_{-\va} +
\delta~T_{-\vb}\right)  \nonumber \\
q^{2J_{3}} = T_{\vb -\va}~,~~~~~~~~q^{-2J_{3}} = T_{\va -\vb}
{}~~~~~~~~~~~~~~~~~~~~~~~
\eq
 with
\bq
q =   \exp\left(+i\frac{B}{2}(\va\times\vb)\right)
\label{q}
\eq
 and  calculating the commutation relations for
  the $J_{\pm}$ and $J_{3}$ using (\ref{mtranslation}) one can
 easily reproduce  (\ref{qa}) if $\alpha \delta = \beta \gamma = -1$.

In the end of this section let us  note that
 that action of \qa  generators $J_{\pm}$ and $J_{3}$ on the wave functions
 on the first Landau level (generalization on the  case of arbitrary
 level $n$ is straightforward)  depends (even after fixing
  $\alpha, \beta, \gamma, \delta$)  both on choice of a fundamental
 cell $(\va,\vb)$ and a basis of the wavefunctions. We shall present here
 the action of \qa on basic wave functions  (\ref{momentumwf}) for  a
 generic  $(\va,\vb)$ for $\alpha = \gamma =1,~
 \beta = \delta = -1$.  Using (\ref{pTaction})  one can find
\bq
J_{+}|p> & = & \frac{
\exp\left(+iBpa_{1}-\frac{i}{2}Ba_{1}a_{2}\right)|p-a_{2}>-
\exp\left(+iBpb_{1}-\frac{i}{2}Bb_{1}b_{2}\right)|p-b_{2}>}
{q - q^{-1}}
 \nonumber \\
q^{\pm 2J_{3}}|p> &= & \exp\left(\pm iBp(b_{1}- a_{1}) -
\frac{i}{2}B (a_{1}-b_{1})(a_{2}-b_{2})\right) |p \pm(a_{2}-b_{2})>
{}~~~~~ \\
J_{-}|p> & = & \frac{
\exp\left(-iBp a_{1} - \frac{i}{2}B a_{1}a_{2}\right) |p + a_{2}> -
\exp\left(-iBp b_{1} - \frac{i}{2}B b_{1}b_{2}\right) |p + b_{2}>}
{q - q^{-1}}
 \nonumber
\eq

Acting many  times by operators $J_{\pm}$
  and $q^{\pm 2J_{3}}$  on state $|p>$
 in  general case when $a_{2}$ and $b_{2}$ are two incommensurable
 real numbers,
  one can obtain the state $|p'>$ with $p' = p \pm n_{1}a_{2}
 \pm n_{2}b_{2}$ arbitrary close to $p$ for large enough
 $n_{1}$ and $n_{2}$. It is possible however to choose the
 fundamental cell in a  more restrictive way. Choosing, for example,
 $a_{2} = b_{2} = b$ we see that $J_{\pm}$  acting on $|p>$
  create  one, not two as in general case, state $|p \mp b>$
 and  $q^{\pm J_{3}}$ are now  diagonal. The natural choice is
  to take $a_{1} = -b_{1} = a$, i.e. to have
\bq
\vec{a} = (a,b),~~~~~~~\vec{b} = (-a,b),~~~~~~~
q = \exp(iBab)
\eq
In this case we get very simple  representation
\bq
J_{+}|p> = \left[\frac{p}{b} - \frac{1}{2}\right]_{q}|p - b>,~~
J_{-}|p> = -\left[\frac{p}{b} + \frac{1}{2}\right]_{q}|p + b>, ~~
q^{\pm J_{3}}|p> = q^{\mp\frac{p}{b}}|p>~~~~
\label{prepresentation}
\eq
 where the quantum symbol $[x]_{q}$ is defined as
\bq
[x]_{q} = \frac{q^{x} - q^{-x}}{q - q^{-1}}
\eq

 Let us calculate the value of the $q$-dimension of the representation
  which is defined as the value of the $q$-deformed Casimir operator
\bq
C_{q} = J_{-}J_{+} + \left[J_{3} + \frac{1}{2}
\right]_{q}^{2} =
        J_{+}J_{-} + \left[J_{3} - \frac{1}{2}
\right]_{q}^{2}
\label{qcasimir}
\eq
When acting on the highest weight vector $|j>$, such that
 $J_{+}|j> = 0,~J_{3}|j> = j|j>$,
 one gets $C_{q}|j>=[j+1/2]_{q}^{2}|j>$ which gives us the
 $q$-dimension of representation. However  representation
(\ref{prepresentation}) is neither of highest nor of lowest weight
 and calculating its  $q$-dimension
\bq
C_{q}|p> = -\left[\frac{p}{b} - \frac{1}{2}\right]_{q}^{2}|p>
+ \left[ \frac{1}{2} -\frac{p}{b}\right]_{q}^{2}|p> = 0
\eq
we get zero.

 Let us note that one  gets  the zero $q$-dimension
 for highest weight representation $|j>$
  with  $j =  (nk - 1)/2$ for $q$ being root of unity
 $q^{k} =  1$ because $[j+1/2]_{q} = [nk/2]_{q} = 0$ in this case.
  The  representation theory for
 $q$ being root of unity was considered in \cite{root}, see
  also \cite{ps}. There are two types of representations:
\begin{itemize}
\item Type-I representations which have $q$-dimension zero
  and are either mixed, i.e. not  highest weight representations,
 or irreducible highest weight representations with
  $j =  (nk - 1)/2$.
\item Type-II representations with nonzero $q$-dimension
  which are
irreducible highest weight representations with $0\leq
 j < k/2 -1$.
\end{itemize}

 In our case the  deformation parameter $q$ is
  arbitrary and depends on our choice
 of fundamental  cell, to be more precise it depends
 on the flux $\Phi = 2Bab$ through this cell
 $q = \exp (i\Phi/2)$. For $q^{k} = 1$ one must have
 $ Bab = 2\pi/k$.
 In the case of Azbel-Hofstadter problem
 which was considered in \cite{wz}, \cite{fk}
  the fundamental cell is the minimal  plaquette
 on the lattice and the interesting case is when the flux through plaquette
 is rational.  Here we shall consider two other problems -
 $2+1$ abelian TMGT  and QHE where the choice of q  also will be
 dictated by  rational numerical parameters of the  corresponding
 problem - the value of the numerical coefficient in front of
 Chern-Simons term or the filling factor in quantum  Hall system.
 One can see that the special values
 of $j$ corresponds to $p = (nk/2 - 1/2)b$. Remembering
 that momentum in $x_{1}$ direction (see(\ref{momentumwf}))
 is $Bp = (nk/2 - 1/2)bB = (2\pi/a)(n/2 - 1/2k)$ we get
 discrete momenta. There is a natural appearence of
 discrete momenta in Landau
 problem  on cylinder or torus and as we shall see in the next sections
 these are precisely the cases which we  shall  be interested in.

\newsubsection{\qa from  a central extension of
 the FFZ trigonometric algebra}

 It is known that a classical  algebra of the
 area-preserving diffeomorphisms on a torus has
    a central extension (which does
  not exist
 in the case of the area-preserving diffeomorphisms on a sphere)
\cite{membrane},\cite{ffz}
\bq
[w_{\vec{n}},w_{\vec{m}}] = (\vec{n}\times\vec{m})~w_{\vec{m}+\vec{n}}
 +  \vec{a}\vec{n}\delta_{\vec{m} + \vec{n},0}
\eq
 as well as a trigonometric FFZ algebra
 \bq
 [ {\cal W}_{\vec{n}},{\cal W}_{\vec{m}}] =
 -2i\sin\frac{B}{2}(\vec{n}\times\vec{m})
 {\cal W}_{\vec{n}+\vec{m}}  +
\vec{a}\vec{n}\delta_{\vec{m} + \vec{n},0}
\label{W1central}
\eq
where the central element is given by an arbitrary vector $ \vec{A}$.

The algebra (\ref{W1central}) can not be obtained from magnetic
 translations (\ref{mtranslation}) because now
 ${\cal W}_{\vec{n}}$ and ${\cal W}_{-\vec{n}}$ do not commute.
 However we still can construct the quantum algebra \qa from
 trigonometric FFZ algebra with a central extension  $ \vec{a}$
 using the same construction as before
 \bq
J_{+} = \frac{1}{q - q^{-1}} \left(\alpha~{\cal W}_{\vec{n}} +
\beta~ {\cal W}_{\vec{m}}\right), ~~~~
J_{-} =  \frac{1}{q - q^{-1}} \left(\gamma~{\cal W}_{-\vec{n}}
 + \delta~{\cal W}_{-\vec{m}}\right)  \nonumber \\
q^{2J_{3}} = {\cal W}_{\vec{m}-\vec{n}},
{}~~~~~~~~q^{-2J_{3}} = {\cal W}_{\vec{n}-\vec{m}}
{}~~~~~~~
q =   \exp\left(+i\frac{B}{2}(\vec{n}\times \vec{m})\right)
\label{qcentralFFZ}
\eq
 Calculating the commutation relations  using \ref{W1central})
  we get (puting as before $\alpha \delta = \beta \gamma = -1$)
\bq
\left[ J_{+}, J_{-}\right] = \frac{q^{2J_{3}} - q^{-2J_{3}}}{q - q^{-1}}
 + \frac{1}{(q - q^{-1})^{2}}\vec{a}\left(
 \alpha\gamma \vec{n} + \beta\delta \vec{m}\right)
\nonumber \\
\left[q^{-2J_{3}},q^{+2J_{3}}\right] =
 \left[{\cal W}_{\vec{n}-\vec{m}},{\cal W}_{\vec{m}-\vec{n}}\right] =
\vec{a}(\vec{n} - \vec{m}) ~~~~~
\eq
 and a commutation relations between
  $J_{\pm}$ and $J_{3}$  are not affected by a central extension $\vec{a}$.
 To get  the \qa commutation relations (\ref{qa})
  we must  have
\bq
\vec{a}(\vec{n} -  \vec{m}) = 0,~~~~~~
\alpha\gamma  + \beta\delta = 0,~~~~~\alpha \delta = \beta \gamma = -1
\eq
 i.e.  the   vector $\vec{a}$ must be orthogonal  to the
 vector $\vec{n}- \vec{m}$ and  there are  3  constraints (complex) for
 four parameters (complex) $\alpha, \beta, \gamma \delta$
\bq
\gamma = - \frac{1}{\beta},~~~~~ \delta = -\frac{1}{\alpha},~~~~~
 \alpha^{2} + \beta^{2} = 0
\eq
 leaving us with one-parameter family
\bq
\alpha = \rho e^{i\phi},~~~
\beta = \rho e^{i\phi \pm i\pi/2},~~~
\gamma =\rho^{-1} e^{-i\phi \pm i\pi/2},~~~
\delta =\rho^{-1} e^{-i\phi \pm i\pi}
\eq

  We see that in a case of  a nontrivial  central extension
  one has to deal with   more restrictions   when  constructing a quantum
algebra.
 First of all the choice of the fundamental cell $\vec{n},
 \vec{m}$ is no longer arbitrary but the basic vectors  must be choosen
  in a way that
 $\vec{n}- \vec{m}$ is orthogonal to  the  vector $\vec{a}$.
  Besides this the  four  parameters $\alpha, \beta, \gamma \delta$
  are completely determined by one (complex) parameter
 (for example, $\alpha$), contrary to
   a  case  $\vec{a}=0$ with   two  independent parametres
 (for example  $\alpha$ and $ \beta$).

\newsection{Quantum symmetry in $2+1$ gauge theory with
 Chern-Simons term}
  \newsubsection{Canonical quantization of the $2+1$
 TMGT}

Let us consider an  abelian topologically massive gauge theory \cite{tmgt}:
\begin{equation}
S_{U(1)} =
 -\frac{1}{4\gamma}\int \sqrt{-g}g^{\mu\alpha}g^{\nu\beta}F_{\mu\nu}
 F_{\alpha\beta} +
\frac{k}{8\pi}\int \epsilon_{\mu\nu\lambda}
A_{\mu}\partial_{\nu}A_{\lambda}
\label{eq:tmgt}
\end{equation}
To perform canonical quantization let us  chose a
  $A_{0}=0$ gauge.
  Representing vector-potential on a plane as
$
A_{i} = \partial_{i}\xi + \epsilon_{ij}\partial_{j}\chi
$
and substituting this decomposition into constraint
\begin{equation}
\frac{1}{\gamma}\partial_{i}\dot{A}_{i} +
\frac{k}{4\pi} \epsilon_{ij}F_{ij} =
0, \label{eq:constraint}
\end{equation}
one  gets
 $\partial^{2} \dot{\xi} = (k\gamma/2\pi)\partial^{2} \chi $.
Neglecting all possible zero modes we  put $\dot{\xi} =
 (k\gamma/2\pi) \chi = (M/2)\chi$. Substituting this constraint into
action (\ref{eq:tmgt}) one gets
\begin{eqnarray}
S = \frac{1}{2\gamma} \int(\partial_{i}\dot{\chi})^{2} -
(\partial^{2}\chi)^{2} - M^{2}\chi\partial^{2}\chi
\end{eqnarray}
which becomes a free  action
   for the  field $\Phi =
\sqrt{\partial^{2}/ \gamma} \chi$
\begin{eqnarray}
S = \frac{1}{2} \int \dot{\Phi}^{2} - (\partial_{i}\Phi)^{2} -
M^{2}\Phi^{2}
\end{eqnarray}
describing the free particle with mass $M =\gamma k/4\pi$.
 In obtaining  this action  we  used
 the  constraint
(\ref{eq:constraint}). However  there are some
 field configurations which are escaped from this constraint.
 It is easy to see  that on the plane   the
 spatial independent  fields $A_{i}(x,t) = {\bf A}_{i}(t)$
 are not affected
 by  (\ref{eq:constraint}) - because both terms
  $F_{ij}$ and  $\partial_{i}E_{i}$ are zero for
 space-independent vector potential (but not electric field
 $E_{i} = \dot{A}_{i}$).
  For these fields  one  gets the
Landau  Lagrangian  (\ref{Laction}) \cite{km}
\begin{equation}
L = \frac{1}{2\gamma}\dot{{\bf A}}_{i}^{2} -
 \frac{k}{8\pi}\epsilon_{ij}{\bf
A}_{i}\dot{{\bf A}}_{j}  \label{eq:Landau}
\end{equation}
which describes the particle with mass $m = \gamma^{-1}$
  on the plane ${\bf A}_{1},{\bf A}_{2}$ in
 a magnetic field $B = k/4\pi$. From (\ref{Lhamiltonian}) the mass gap is
  $M = B/m = \gamma k/4\pi$   which is precisely the  mass of gauge
 particle.

Let us note that  ${\bf A}_{1}$ and ${\bf A}_{2}$
 belong to  the configuration
 space, however if reduced to  the first Landau level the
 configuration space is transformed into the phase space as we
 have discussed before. In this case a reduction to the
 first Landau level means $ m = 1/\gamma \ra 0$, i.e. the
 reduction to the  pure Chern-Simons theory which
 is an exactly solvable  $2+1$ dimensional topological field theory.

Is it possible to consider
 a  constant gauge field  as a physical, i.e. gauge invariant
  variable  in the theory ? Can one simply gauge away the constant field ?
 To answer this question we must define the boundary conditions at
 infinity, i.e. to compactify our plane into a 2-dimensional Riemann
  surface of genus $g$.
  It is well-known  that any  one-form $A$ can be uniquely
 decomposed  according to  Hodge  theorem as
\begin{eqnarray}
 A = d\xi + \delta\chi + {\bf A},~~~~~~~~~~~
d{\bf A} = \delta{\bf A} = 0
\end{eqnarray}
which generalizes the decomposition on the plane we have used
 before. The harmonic form
${\bf A}$ equals
\begin{eqnarray}
{\bf A} = \sum_{p=1}^{g} ({\bf A}^{p}\alpha_{p} + {\bf B}^{p}\beta_{p})
\label{1forms}
\end{eqnarray}
where $\alpha_{p}$ and $\beta_{p}$ are canonical harmonic 1-forms
 ($1$-cohomology) on a
Riemann surface and  there
 are precisely $2g$ harmonic 1-forms  on genus $g$ Riemann surface
(two in  case of a torus which are these two constant modes
 we have discussed).	After diagonalization
  one finds that there are $g$ copies of the Landau problem and
  the  total Hilbert space  ${\cal H}$
  of the abelian topologically massive gauge
theory
\bq
{\cal H} = {\cal H}_{\Phi}\otimes\prod_{i=1}^{g}
{\cal H}_{{\bf A}}
\eq
is the product of the free massive particle  Hilbert space
$ {\cal H}_{\Phi}$  and $g$ copies
of the  Landau problem's  Hilbert space ${\cal H}_{{\bf A}}$.

 There is a  dependence
 on the moduli of the Riemann surface due to   the
  dependence of the $F^{2}$  term
 in (\ref{eq:tmgt})
 on metric $g_{\mu\nu}$. It is easy to see that
 for $g_{00} = 1$ and $g_{ij}= \rho(x) h_{ij} (\tau)$ the
  $F_{0i}^{2}$ term does not depend  on  conformal factor
 $\rho$.  Let us  consider dependence on the moduli $\tau$ in the
 most simple case of a  torus where $\tau$ is a complex number  and
 metric $h_{ij}$ can be parametrized as
\begin{eqnarray}
 h^{ij} = \frac{1}{(Im\tau)^{2}} \left( \begin{array}{cc}
1 &
Re\, \tau\\
 & \\
Re \, \tau & |\tau|^{2} \\ \end{array}\right)  \;\;\;\;\;
h_{ij} = \left( \begin{array}{cc}
|\tau|^{2}&
-Re\, \tau\\
 & \\
-Re \, \tau & 1 \\ \end{array}\right) \label{h}
\end{eqnarray}
 and  $h = det h_{ij} = (Im\tau)^{2}$.
  Lagrangian takes the form
 \begin{equation}
L = \frac{1}{2\gamma}\sqrt{h}h^{ij}\dot{{\bf A}}_{i}
\dot{{\bf A}}_{j}  - \frac{k}{8\pi}\epsilon^{ij}{\bf
A}_{i}\dot{{\bf A}_{j}}  \label{landau}
\end{equation}
which can be transformed to diagonal form  (\ref{eq:Landau})
 for new fields
\bq
{\bf A}_{(a)} = e^{i}_{(a)}{\bf A}_{i}
\eq
 where zweibein $e^{i}_{(a)}$ defines the metric
$h^{ij} = e^{i}_{(a)}e^{j}_{(b)} \delta^{(a)(b)}$
 and  $\epsilon^{(a)(b)} e^{i}_{(a)}e^{j}_{(b)} \sim\epsilon^{ij}$
  It is easy to  find that
\bq
 \left(\begin{array}{c}
 {\bf A}_{(1)}\\{\bf A}_{(2)} \\ \end{array}\right)
  = \left( \begin{array}{cc}
 1 &  Re\, \tau \\
0 &  Im\, \tau \\ \end{array}\right)
\left(\begin{array}{c}
 {\bf A}_{1}\\{\bf A}_{2} \\ \end{array}\right)
\label{new}
  \eq
   In terms of the new variables the Lagrangian
 (\ref{landau}) takes the form
 \begin{equation}
L = \frac{1}{2\gamma~ Im \,\tau }\dot{{\bf A}}_{(i)}^{2}
  - \frac{k }{8\pi ~Im \,\tau}\epsilon^{(i)(j)}{\bf
A}_{(i)}\dot{{\bf A}}_{(j)}  \label{diaglandau}
\end{equation}
and we see that  the Chern-Simons coefficient depends on
 moduli: $k \rightarrow k/Im\,\tau$. However the mass gap
 is unchanged  because $\gamma$ is also
 changed $\gamma  \rightarrow\gamma~
 Im\,\tau$
 and  $M = \gamma k/4\pi$ does not depend
 on $\tau$.

 Thus we get the Landay problem on the plane $ ({\bf A}_{(1)},
 {\bf A}_{(2)})$.
   However we forgot about  large  gauge transformations acting
  on the
quantum-mechanical coordinates  ${\bf A}_{i}\rightarrow{\bf A}_{i}+
 2\pi\, N_{i}$,
 where $N_{i}$ are integers. These transformations  act on gauge
 potential because
  the only  gauge-invariant objects one can construct for
 ${\bf A}_{i}$  - Wilson lines
 \bq
W(C) = exp(i\oint_{C} A_{\mu} dx^{\mu})
\eq
 are invariant
 under these transformations (we choose coordinate on a torus in a
 way  that $x^{1} \sim x^{1}+1$ and $x^{2} \sim x^{2} + 1$)
 and one can consider torus  $0\leq{\bf A}_{i}<2\pi$
  with the area $(2\pi)^{2}$.   However
 after we  consider the new variables   ${\bf A}_{(i)}$
 one gets the torus (see (\ref{new}))
  generated by the shifts $2\pi$ and $2\pi\tau$
  with  an area $S = (2\pi)^{2}~Im \tau$.

  Let us note that being reduced
  to the first Landau level
  this torus becomes the phase space - thus for the consistent quantization
  this area must be proportional to the  integer
  (the   total number of the states must be integer).
   It is known
 that the density of states $\rho$  on Landau level equals
 to
 $B/2\pi$, where $B$ is a magnetic field. In our case the
 "magnetic field" in   $({\bf A}_{(1)}, {\bf A}_{(2)})$  plane can be
 easily obtained from (\ref{diaglandau}) and equals to
 $B = (k/4\pi)~ Im\,\tau$, thus the  total number of states
 will be $N = (1/2\pi) (k/4\pi~ Im\,\tau) \times (2\pi)^{2}~Im\,\tau =
 k/2$.  and does not depend on $\tau$  but only on $k$.

  We see that it is  possible to
  factorize over whole large gauge transformations only for even
 $k$, for rational $k = 2m/n$ one can not any longer maintain
  the whole large gauge transformations group and only the
 subgroup with  shifts
 ${\bf A}_{i}\rightarrow{\bf A}_{i}+n N_{i}$,
  are  survive and so
   one gets the torus in a phase space  $0\leq{\bf A}_{i}
 < 2\pi n$  with the total area  $(2\pi n)^{2}$ (we put
 $\tau = i$ here because as we have mentioned before the
 number of states does not depend on moduli)  and the number of states
  is $N = (1/2\pi) ( m/2\pi n) \times (2\pi n)^{2}  =
  mn$.

\newsubsection{\qa in Chern-Simons theory and
 Landau problem on a torus}

 Now it is clear that to study the properties of the
 ground state in TMGT (or the whole Hilbert space in a
 topological Chern-Simons theory) one has to consider a
 Landau problem on a torus and this problem was considered
 in \cite{dnhr}.  Let us start from the
 first Landau level  wavefunction on a
 plane $({\bf A}_1,{\bf A}_2)$
\bq
\Psi({\bf A}_1,{\bf A}_2) =
\exp \left( -\frac{ik}{8\pi}{\bf A}_1{\bf A}_2+
\frac{ik}{4\pi} p {\bf A}_1
 - \frac{k}{8\pi} ({\bf A}_2 - p)^2  \right) = \nonumber \\
 \exp\left( -\frac{k}{16\pi} {\bf A} \bar{{\bf A}} \right)
 \exp\left( -\frac{k}{8\pi} p^2 +\frac{k}{16\pi} \bar{{\bf A}}^{2}
 +  i\frac{k p}{ 4\pi} \bar{{\bf A}}\right)
\label{psi}
\eq
 which can be easily obtained from $ (\ref{momentumwf}) $
 substituting $B = k/4\pi$ and using ${\bf A}_1,{\bf A}_2$
 notation instead of $x_{1}, x_{2}$ and ${\bf A}( \bar{{\bf A}})$
 instead of $z\bar{z}$.

 To get the correct wave functions  on a torus
 let us consider  the  simplest case
  $\tau = i$.  One can construct  a torus first  making a periodicity
 in ${\bf A}_1$ direction with period $2\pi$,
 which leads to an evident quantization codition
   $p_{n} =  4\pi n /k$. At the same time $p$ is a ${\bf A}_2$
 coordinate of the center of the wavepacket and for  first
 $k/2$ numbers $p_{n}$ (for even $k$)   this coordinate is
 inside an interval
 ${\bf A}_2 \in [0, 2\pi)$\footnote{~ In a general  case
 of rational  $k = 2m/l$ it will be first $lm$
 numbers $p_{n} = 4\pi n/ kl$  which gives us the   coordinate of the
 wavepacket center in the interval ${\bf A}_2 \in [0, 2\pi l)$}.
 Now to make  a torus, i.e. make a periodicity
  in  ${\bf A}_2$ direction with period $2\pi$ (we consider now only
 even $k$) one has to sum  over all $p = 2\pi n$ and it is easy to
  see that for even
 $k$ there are $k/2$ different classes $p = 4\pi r/k + 2\pi n$,~
 $r = 1,\ldots k/2;~ n \in Z$ which gives $k/2$ basic wave functions.

\bq
\Psi_{1}(r|{\bf A}_1,{\bf A}_2) =
\exp\left(-\frac{k}{16\pi} {\bf A}\bar{{\bf A}}\right)
{}~~ \exp\left(\frac{k}{16\pi } \bar{{\bf A}}^{2} \right)~~
 \theta\left[ \begin{array}{c} 2r/k \\
 0\end{array} \right] \bigl(\frac{ k \bar{{\bf A}}}
{4\pi}|\frac{i k }{2})  = \nonumber \\
  \exp\left(-\frac{k}{16\pi} {\bf A}\bar{{\bf A}}\right)
\exp\left( \frac{k}{16\pi} \bar{{\bf A}}^{2}\right)
\sum_{n}
 \exp \left[-\frac{\pi k}{2} (n + 2r/k)^{2}  +
i \frac{k \bar{{\bf A}}}{2}(n + 2r/k)\right]
\label{Psi}
  \eq
where $r = 1,2,\ldots k/2$ and  theta function  is  defined as
 follows
\bq
\theta\left[ \begin{array}{c} \alpha \\
 \beta \end{array} \right] \bigl(z|\tau) =
 \sum_{n}
 \exp[ i\pi\tau (n+\alpha)^{2} +
 2\pi i (n+\alpha)(z+\beta) ]
\label{theta}
\eq

Let us use the compact notation $||r>$ for the wave function
$ \Psi_{1}(r|{\bf A}_1,{\bf A}_2)$ on a torus with $\tau = i$.
   Using (\ref{Psi}) and
 (\ref{momentumwf})  we can write
\bq
||r> = \sum_{m= -\infty}^{\infty} |2\pi(m + 2r/k)>, ~~~~~~~~~~
||r> \equiv ||r + k/2>
\label{Psi1}
\eq
 Now let us consider the action of the \w  (or magnetic translations)
 generators (\ref{W1}), (\ref{mtranslation})
 on the wave functions  (\ref{Psi1}).    It is easy to see that
  in this case one is dealing with generators
  $T_{\vec{\xi}}$,  where
$\vec{\xi} = \vec{n}/B =
 \frac{4\pi}{k}\vec{n}$ and $n_{1}, n_{2}$ are integers.
 Using equation (\ref{pTaction})  we get from (\ref{Psi1})
\bq
T_{\vec{\xi}}~||r> = \sum_{m=-\infty}^{\infty}
{}~ T_{\vec{\xi}}~ |2\pi(m + 2r/k)> =
  \exp\left(-\frac{2\pi i}{k} n_{1}
 (n_{2} - 2r)\right) ||r-n_{2}> ~~~~
\label{representation}
\eq

  This is a   demonstration  that the ground state (first Landau
 level) wave functions in the topologically massive
 gauge theory (not only in the pure Chern-Simons case
 with infinite mass gap)  form a unitary representation
  of a quantum \w, i.e. FFZ algebra
 (\ref{W1}). Let us note that states at higher Landau levels
 which can be obtained from  ground state
  wave functions $\Psi_{r}$ by the
 action of the $a^{+}$ operator (\ref{La})
\bq
\Psi_{N}(r|{\bf A}_1,{\bf A}_2)
 = \frac{(a^{+})^{N}}{\sqrt{N!}}\Psi_{1}(r|{\bf A}_1,{\bf A}_2)
\eq
 form unitary equivalent representations  because generators
  of \w (magnetic translations)  are built from $b$ and $b^{+}$
 operators only and thus commute with $a$ and $a^{+}$.

Now let us consider the "minimal" quantum algebra \qa  with generators
\footnote{This is only one possible choice and one can consider
 another constructions, choosing, for example, not $T_{\pm 1, \pm 1}$
 but $T_{\pm n_{1}, \pm n_{2}}$. In  that case the generators
 $J_{\pm}$ will shift state $||r>$ to $||r \mp n_{2}>$}
\bq
J_{+} = \frac{1}{q-q^{-1}}\left(T_{(1,1)} - T_{(-1,1)}\right),~~~~~~~~~~
J_{-} = \frac{1}{q-q^{-1}}\left(T_{(-1,-1)} - T_{(1,-1)}\right) \nonumber \\
q^{2J_{3}} = T_{(-2,0)},~~~~~~~~q^{-2J_{3}} = T_{(2,0)}, ~~~~~~~~~~~~~~
q = \exp\left(\frac{4\pi i}{k}\right)
\label{torusqa}
\eq
where notation $(n_{1},n_{2})$ corresponds to a vector
 $\vec{\xi} = (\xi_{2}, \xi_{2}) = \frac{4\pi}{k}(n_{1},n_{2})$.
 These generators act on  states (\ref{Psi1}) in the following way:
\bq
J_{+}||r> & = &~~ [r-1/2]_{q} ||r-1> \nonumber \\
J_{-}||r> & = &- [r+1/2]_{q} ||r+1> \\
q^{\pm 2J_{3}}||r> & = & q^{\mp 2r} ||r> \nonumber
\eq

 We have two different types of representations
  in  case of $k = 4n$ and $k = 4n + 2$
 ( don't forget that here we are dealing only with even $k$).

In the case when $k = 4n + 2$ we have a highest and a lowest weight
 vectors. In this case  $q = \exp\left(\frac{2\pi i}{2n + 1}\right)$
  and it is easy to see that $[n + 1/2]_{q} = 0$.  This means that
 $||n+1>$ is the highest and $||n>$ is the lowest weight vectors
\bq
J_{+}||n+1> & = &~~ [n + 1/2]_{q} ||n~~~> ~~ = ~0 \nonumber \\
J_{-}||~n~>  & = & - [n + 1/2]_{q} ||n+1> =~ 0
\eq
 and we have $2n+1$-dimensional representation
\bq
||n+1>
\stackrel{J_{-}}{\longrightarrow} \ldots
\stackrel{J_{-}}{\longrightarrow} ||2n+1>
\stackrel{J_{-}}{\longrightarrow} ||1>
\stackrel{J_{-}}{\longrightarrow} ||2>
\stackrel{J_{-}}{\longrightarrow} \ldots
\stackrel{J_{-}}{\longrightarrow}||n> \nonumber \\
||n+1>
\stackrel{J_{+}}{\longleftarrow} \ldots
\stackrel{J_{+}}{\longleftarrow} ||2n+1>
\stackrel{J_{+}}{\longleftarrow} ||1>
\stackrel{J_{+}}{\longleftarrow} ||2>
\stackrel{J_{+}}{\longleftarrow} \ldots
\stackrel{J_{+}}{\longleftarrow}||n>
\label{4n+2}
\eq

 In  the case $k = 4n$ there are no highest and/or lowest weight vectors
  and we have  cyclic representation with dimension $k/2 = 2n$
\bq
\ldots
\stackrel{J_{-}}{\longrightarrow} ||1>
\stackrel{J_{-}}{\longrightarrow} ||2>
\stackrel{J_{-}}{\longrightarrow} \ldots
\stackrel{J_{-}}{\longrightarrow} ||k/2 = 2n>
\stackrel{J_{-}}{\longrightarrow} ||1>
\stackrel{J_{-}}{\longrightarrow} \ldots
\label{4n}
\eq
  and the the same (with opposite directed arrows) for $J_{+}$.
The $q$-dimension in both cases is zero.

 Let us note  that one can get the highest  weight representation
 in  the case $k = 4n$ if instead of usual periodical boundary conditions
 on a torus we shall consider the twisted boundary conditions in
${\bf A}_{1}$ direction which leads to modified quantization condition
   $p_{n} =  4\pi (n-\alpha) /k$ where $\alpha \in [0,1)$ defines
  an  additional phase factor (twisting) $\exp(2\pi i \alpha)$
  which  arises  in a wave function on cylinder (and torus) after $2\pi$
 shift   in ${\bf A}_{1}$ direction. One can get this shift if there is
 a flux through cylinder (or torus) $\Phi = 2\pi \alpha$. In that case
 the $k/2 = 2n$ state vectors will be $|| r-\alpha>,~~~r = 1,\ldots, 2n$
 and one can have  highest  and lowest  weight vectors for $k = 4n$
  for $\alpha = 1/2$. This corresponds to antiperiodic boundary conditions
 in ${\bf A}_{1}$ direction or to a flux $\Phi = \pi$.   Then it is easy to
  that
 $||n+1/2>$ is the highest and $||n-1/2>$ is the lowest weight vectors
\bq
J_{+}||n+1/2> & = &~~ [n]_{q} ||n-1/2>~ = ~0 \nonumber \\
J_{-}||n-1/2> & = & - [n]_{q} ||n+1/2>~ = ~0
\eq
because here $q = \exp(\pi i/n)$
  and we have $2n$-dimensional representation  completely analagous
 to (\ref{4n+2}).

One can consider in a  same way the case of rational $k = 2m/n$.
 Let us note that there is an  ultimate connection between $2+1$
 topological Chern-Simons theory and $1+1$ conformal field theory
  (CFT)  \cite{witten}. In  our case  the corresponding
 conformal field theory is $c=1$ model and the states on the
 first Landau level are in one-to-one correspondence
 with the conformal blocks of a $c=1$ model \cite{confblo}.
  This means that there is \qa symmetry in $c=1$ CFT  and
 conformal blocks are  a representation of quantum algebra.

\newsection{Quantum symmetry in a quantum Hall system}
\newsubsection{Area-preserving diffeomorphisms and
 incompressibility in a quantum Hall system}

The area-preserving diffeomorphisms and corresponding
\w symmetry were  discussed recently in quantum Hall systems
 in \cite{sakita}, \cite{ctz}. The   physical reason  for
  the very existence of this symmetry  is based on
  a  Laughlin idea \cite{laughlin}
 that ground state  of the quantum Hall system at rational
 (integer and fractional) values of a
 filling factor $\nu = 2\pi\rho/B$, where $\rho$ is an  electron density,
 is described by  an  incompressible quantum fluid, i.e.
  there is an energy  gap in a spectrum of  excitations.
 In the case of an integer quantum Hall effect (IQHE) this
 gap is a mobility gap between Landau levels (taking
 into account the disorder and localised states)
  which  for strong
 magnetic field $B$ is much larger then the energy of
 Coulomb repulsion between electrons. Thus one can neglect
 the interaction in the IQHE effect and consider it as a
 system of $\nu$ completely filled Landau levels.
 The fractional quantum Hall effect (FQHE) occurs in low-disorder,
  high-mobility samples with partially filled Landau levels.
 In this case there is no single-particle gap and  only
 after takking into account many-body correlation due
 to the Coulomb repulsion the excitation gap appears
  in a spectrum as a collective effect. In the case of filling
 $\nu = 1/(2p +1)$  the ground state wave function is described
 by Laughlin wave function \cite{laughlin} (let us note that in
 our notation the wave function  on the first Landau level  depends
 on $\bar{z}$, not $z$)
\bq
\Psi(z_{1},\bar{z}_{1},\ldots,z_{n},\bar{z}_{n})
  = \prod_{i<j}(\bar{z}_{i} -
 \bar{z}_{j})^{2p+1} \exp\left(-\frac{B}{4}\sum_{i}|z_{i}|^{2}\right)
\label{laughlin}
\eq
In the case $p = 0$ this function describes a completely filled
 first Landau level.

Now  let us consider an   operator
 \bq
{\cal \bf L}_{n,m} = \sum_{i=1}^{N} {\cal  L}^{i}_{n,m}
= \sum_{i=1}^{N}(b^{+}_{i})^{n+1}b_{i}^{m+1},~~~~~
n,m \geq -1
\eq
which is the sum of $N$ independent operators (\ref{Lnm}).
 It is easy to see that the commutation relations for these
 operators are the same as for one-particle ones
${\cal  L}_{n,m}$ (\ref{comLnm}). If we are on a first Landau level
 the angular momentum  is given by
 $\sum_{i=1}^{N}(b^{+}_{i})b_{i}= {\cal \bf L}_{0,0}$ and one can see that
 $[{\cal \bf L}_{0,0},{\cal \bf L}_{n,m}]=(n-m){\cal \bf L}_{n,m}$, i.e.
 ${\cal \bf L}_{n,m}$  with $n < m$  are decreasing an  angular momentum and
   in result
 compress the Hall liquid. Thus, being applied to the uncompressible
  completely filled   level  it must  annihilate it
\bq
{\cal \bf L}_{n,m} \Psi_{\nu = 1} = 0,~~~~~ n < m
\label{LPsi}
\eq
There exists a second-quantized  representation for these
 generators
\bq
{\cal \bf L}_{n,m} = \int d^{2}x \hat{\Psi}^{\dagger}(\vec{x},t)
(b^{+})^{n+1}b^{m+1} \hat{\Psi}(\vec{x},t)
\eq
where  $\hat{\Psi}(\vec{x},t)$ is the field operator for the
 fermions in an external magnetic field
\bq
\hat{\Psi}(\vec{x},t) = \sum_{n=1}^{\infty} \sum_{k}
 F_{k}^{(n)} \phi^{(n)}_{k}(\vec{x}) \exp\left(
-i\frac{B}{m}(n+\frac{1}{2})\right)
\eq
 and $\phi^{(n)}_{k}(\vec{x})$ are the wave functions  on the
 $n$-th Landau level and $ F_{k}^{(n)} (F_{k}^{(n)~\dagger}$ are
 the fermionic creation and annihilation operators.
   One can use this
 representation to obtain the expression for ${\cal \bf L}_{n,m}$
 in terms of fermionic creation and annihilation operators
 (see details in \cite{ctz}) and then  it is
 easy to show that conditions (\ref{LPsi}) are valid for
 arbitrary integer-valued filling $\nu$ also. The case of
 the fractional filling was also considered in \cite{ctz}
 and in a recent paper \cite{fv}. One can also show
 using the second-quantized
  representation and (\ref{WLconnection})
 that  the Fourier-transformed second-quantized
 density operator $\rho(\vec{k}) =
 \int d^{2}x \exp(i\vec{k}\vec{x}) \hat{\Psi}^{\dagger}(x)
 \hat{\Psi}(x)$, being projected on the first Landau level,
 becomes  proportional to a  ${\cal W}_{k,\bar{k}}$ generator
 with $k(\bar{k}) = k_{1}\pm k_{2}$.

  The incompressibility of the  quantum Hall liquid
 thus naturally leads to a some  \w symmetry (to be more
 precise it is called $W_{1+\infty}$ in a  literature).
 If one considers a
 droplet of a quantum Hall liquid it is evident
 that the only possible deformations of  this  droplet
  preserving  the area turns are the waves  at the boundary of the
 droplet describing the deformation of shape - the
 so-called edge excitations \cite{edge}.

\newsubsection{ Magnetic translations and \qa in
 quantum  Hall system}

After  this brief review of  area-preserving diffeomorphisms
 and \w symmetries in  a  quantum Hall system let us
  consider our construction of \qa and ask the obvious
 question - what will be a  natural value for a deformation
  a  parameter $q$ and  how this symmetry will act on the
 physical states.  Let us note that there are two possibilities
 to construct the  generators of a \w algebra in this case.
 First of all we can use the  connection (\ref{WLconnection})
 between ${\cal W}_{n,\bar{n}}$ and ${\cal  L}_{n,m}$
 and construct ${\cal \bf W}_{n,\bar{n}}$ generators in a
 complete analogy with ${\cal  \bf L}_{n,m}$, i.e. summing
 over all one-particle generators ${\cal W}^{i}_{n,\bar{n}}$
\bq
{\cal \bf W}_{n,\bar{n}} =
\sum_{i =1}^{N}{\cal W}^{i}_{n,\bar{n}} =
\sum_{i =1}^{N}\exp\left(\frac{1}{2}(n b^{+}_{i} - \bar{n} b_{i})\right)
\eq
One can rewrite it in a second-quantized form as
\bq
{\cal \bf W}_{n,\bar{n}} =
 \int d^{2}x \hat{\Psi}^{\dagger}(\vec{x},t)
\exp\left(\frac{1}{2}(n b^{+} - \bar{n}b)\right) \hat{\Psi}(\vec{x},t)
 =  \nonumber \\
\exp\left(\frac{B}{4}n\bar{n}\right)
  \int d^{2}x \hat{\Psi}^{\dagger}(\vec{x},t)
\exp\left(-\frac{1}{2} \bar{n}b\right)
\exp\left(\frac{1}{2} nb^{+}\right) \hat{\Psi}(\vec{x},t)
\eq
It is easy to see that being projected on the first Landau level
  one can effectively substitute $ b^{+} =i B\bar{z},~~
 b =  -i B z$  (see (\ref{aabb})) and thus get  the
 Fourier-transformation of the density operator
 $\rho(\vec{x}) = \hat{\Psi}^{\dagger}(\vec{x},t)\hat{\Psi}(\vec{x},t)$
 projected on the first Landau level
\bq
{\cal \bf W}_{\vec{n}} =
 \exp\left(\frac{B}{4}\vec{n}^{2}\right)
  \int d^{2}x \hat{\Psi}^{\dagger}(\vec{x},t)
\exp\left(i\frac{B}{2}(n\bar{z} +  \bar{n} z)\right)
 \hat{\Psi}(\vec{x},t) = \nonumber \\
 \exp\left(\frac{B}{4}n\bar{n}\right)
  \int d^{2}x~ \exp(iB\vec{n}\vec{x})~~\rho(\vec{x})
{}~~~~~~~~~~~~~~~
\eq
Let us note that because of the projection on a given level
 the time dependence  disappears from  $\rho(\vec{x})$ because
 all $\exp(-i(B/2m) t)$ factors from $\hat{\Psi}(\vec{x},t)$
 will be cancelled by $\exp(+i(B/2m) t)$ factors  in
 $\hat{\Psi}^{\dagger}(\vec{x},t)$.

We can, however, define the total \w generators acting on the
 whole quantum Hall system in another way, namely one can
 multiply all one-particle generators ${\cal W}^{i}_{\vec{n}}$
  or equivalent magnetic translation generators $T_{\vec{n}}$
  and get
\bq
 {\cal T}_{\vec{\xi}} = \prod_{i=1}^{N} T^{i}_{\vec{\xi}};
{}~~~~~~~~
{\cal T}_{\vec{\xi}}{\cal T}_{\vec{\eta}} =
 \exp\left(-i\frac{B}{2} N (\vec{\xi}\times\vec{\eta})\right)
{\cal T}_{\vec{\xi}+ \vec{\eta}}
\label{hallW}
\eq
  Before we shall discuss the quantum algebra structure let us
 consider how the translations (\ref{hallW}) acts on the
 Laughlin wave function (\ref{laughlin}). Using (\ref{Taction})
 and the fact that  $T^{i}_{\vec{\xi}}$ factors act independently
 on $z_{i},\bar{z}_{i}$ arguments, we can get after simple
 calculations
\bq
{\cal T}_{\vec{\xi}}\Psi(z_{1},\bar{z}_{1},\ldots,z_{n}\bar{z}_{n})
  = \exp\left(-\frac{B}{4}\xi\bar{\xi}
  -\frac{B}{2}\xi \sum_{i}\bar{z}_{i} \right)
\Psi(z_{1},\bar{z}_{1},\ldots,z_{n},\bar{z}_{n})
\label{Tlaughlin}
\eq
One can see that up to an overall phase factor
 depending only on the center of mass coordinate
 $\sum_{i}\bar{z}_{i}$ (which is  absent in a center of
 mass frame) the Laughlin wave function is invariant
 under the magnetic translations.

Let us consider how the quasihole wave function
\bq
\Psi(u,\bar{u};z_{1},\bar{z}_{1},\ldots,z_{n},\bar{z}_{n})
  = \prod_{i}(\bar{u} -\bar{z}_{i})
\prod_{i<j}(\bar{z}_{i} - \bar{z}_{j})^{2p+1}
\exp\left(-\frac{B}{4}\sum_{i}|z_{i}|^{2}\right)
\label{quasihole}
\eq
is transformed under the action of ${\cal T}_{\vec{\xi}}$.
 Repeating the same  arguments one gets (let us note that
 $T$ acts only on $z_{i}$, not quasihole coordinate $u$!)
\bq
{\cal T}_{\vec{\xi}}
\Psi(u,\bar{u};z_{1},\bar{z}_{1},\ldots,z_{n}\bar{z}_{n})
= \prod_{i}(\bar{u}-\bar{\xi} -\bar{z}_{i})~
{\cal T}_{\vec{\xi}}
\Psi(u,\bar{u};z_{1},\bar{z}_{1},\ldots,z_{n},\bar{z}_{n})
\eq
and we see that ${\cal T}_{\vec{\xi}}$ shifts a
 quasihole  coordinate $u \ra u - \xi, \bar{u} \ra \bar{u}
 - \bar{\xi}$.

Now let us construct the \qa generators from
  ${\cal T}_{\vec{\xi}}$. Taking as usually
\bq
J_{+} = \frac{1}{q - q^{-1}} \left({\cal T}_{(a,b)} -
{\cal T}_{(-a,b)}\right),
J_{-} =  \frac{1}{q - q^{-1}} \left({\cal T}_{(-a,-b)}
 - {\cal T}_{(a,-b)}\right),
q^{\pm 2J_{3}} = T_{(\mp 2a,0)}
{}~
\eq
 we get the quantum algebra \qa with
\bq
q =   \exp\left(iBNab\right)
\eq
where $N$ is the total number of electrons.
To construct the "minimal" \qa one have to
  choose the minimal $a$ and $b$. It can be shown that
 for a system with sizes $L_{1}$ and $L_{2}$ the minimal
 shifts $a$ and $b$ must be choosen as (see discussion
 in section $3.2$)
\bq
a = \frac{2\pi}{L_{2}B},~~~~ b = \frac{2\pi}{L_{1}B}
\eq
 With this choice one has
\bq
q = \exp\left(iNBab)\right) =
\exp\left(2\pi i \frac{ 2\pi N}{B L_{1}L_{2}}\right) =
 \exp (2\pi i\nu)
\eq
 where   we used the fact that filling factor
$\nu$  is defined as the ratio of total number of particles
 $N$ to the total number of available  states on the
 Landau level $B L_{1}L_{2}/2\pi$, i.e
$\nu = 2\pi N/B L_{1}L_{2}$.
We  see that the ground state is a singlet (up to total
 shift of the center of mass) under the action of a quantum group.
 However the quasihole wave function is transformed under \qa
  and the basis (\ref{quasihole}) is not the convenient one to
  study the action of quantum algebra - one can see that $J_{3}$
   in this basis is not diagonal. One can consider the QHE on a
 cylinder or a torus and then  using the arguments from the
 section $3.2$ we shall get the same representations as
 (\ref{4n+2}) and (\ref{4n}). Let us remember that in a
first case we had $q = \exp(2\pi i/ (2p +1))$\footnote{
 In section $3.2$ we used letter $n$ instead of $p$ here}
  which is precisely
 the case of the  fractional filling factor  with  an
odd denominator
 $\nu = 1/(2n +1)$. It is amusing that in this case
 we had $2p+1$-dimensional representation with  higest
  and lowest weight vectors. In the case
 $q = \exp(2\pi i/ 2p )$ which corresponds to the
fractional filling factor  with  an even denominator
 $\nu = 1/2n$ there were no  higest
  and lowest weight vectors. It is interesting that
 it is Pauli principle which prescribes the odd
 denominators  for Laughlin wave function (\ref{laughlin})
  to be antisymmetric under the permutaions $z_{i}
\longleftrightarrow z_{j}$. It is amusing that our quantum algebra
 construction know about it in some way. Let us also
 remind that if it is a flux $\Phi = \pi$ through the
 cylinder     the highest weight representation
  will be in the even denominator case
 $q = \exp(\pi i/ p )$ as have been discussed
 at the end of section $3.2$. However this flux
 corresponds to the antiperiodic boundary condition
 which may be treated as a statistical transmutations
 from fermions to bosons - the even
 denominators  appearence in this case becomes obvious.

 Let us also mention another  relation between
 quantum algebras in Chern-Simons theory and in a
 quantum Hall system. For CS theory we got the
 deformation parameter $q_{CS} = \exp(4\pi i/k)$  and
 for quantum Hall system $q_{QH} = \exp(2\pi i\nu)$.
 One can describe the large-scale  properties of
  a quantum Hall system by an effective
  Ginzburg-Landau theory with the
  Chern-Simons term  for "statistical" gauge
 field $a_{\mu}$ (don't mix with electromagnetic
 field $A_{\mu}$) \cite{ginzlandauqhall},
 \cite{fkz}. In a simplest case $\nu = 1/(2p+1)$
 one  has the effective  action with a
 Chern-Simons term
 \bq
\frac{1}{4\pi\nu} \epsilon_{\mu\nu\lambda}
a_{\mu}\partial_{\nu}a_{\lambda}
\eq
 and comparing with (\ref{eq:tmgt}) we get
\bq
\nu = \frac{2}{k},~~~~q_{QH} = \exp(2\pi i\nu)
 = \exp\left(\frac{4\pi i}{k}\right) = q_{CS}
\eq
i.e.   for both descriptions - microscopical
 and effective, based on Chern-Simons theory,
 we got the same quantum algebra \qa.

In a more general case $\nu = m/(2pm+1)$
 \cite{jain} instead of
 one abelian Chern-Simons field we have $m$ ones
 \cite{fkz} and  have to generalize  our quantum
 group construction for multidimensional Landau problem.
 This interesting question will be discussed elsewhere.

\newsection{Conclusion}

 We discussed in this paper  how the
 quantum algebra \qa can be constructed from the
 generators of the \w transformations. The physical model
 for this construction was the Landau problem and the
 \w generators were nothing but the magnetic translations.
 However as we had demonstrated in the case of the
 FFZ algebra with a nontrivial central extension
  one can construct the quantum algebra from the
 \w algebra which can not be obtained from the
 group of magnetic translations.  Two examples of a
 quantum \qa algebra were considered  - with
 $q_{CS} = \exp(4\pi i/k)$ in
 an abelian $2+1$ gauge theory with  a
 Chern-Simons term and with $q_{QH} = \exp(2\pi i\nu)$, where
 $\nu$ is a filling factor, in a quantum Hall systems.
 We have demonstrated using the effective large-scale
 description of a quantum Hall system in terms
 of "statistical" Chern-Simons field that
 $q_{CS} = q_{QH}$ in the case of filling factor
 $\nu = 1/(2p+1)$. In  a general case
 $\nu = m/(2pm+1)$  we have to consider  a
  quantum  algebra
 construction for a general $\prod_{i=1}^{m} U_{i}(1)$
 Chern-Simons theory -  which reduces to a   multidimensional
  Landau problem.

 There are  a lot of other interesting questions which are
 still open.  First of all it is interesting to
 generalise this construction to the  nonabelian
 case for the Chern-Simons theory.
 It is unclear what will be the analog of quantum
 algebra in this case. Let us note that this question
 include not only  topologically massive nonabelian
 Yang-Mills theory but also $2+1$-dimensional gravity
 which can be considered as  Chern-Simons gauge theory
 too \cite{gravity}.

It also will be extremely interesting to understand if
 there is a more general connection between area-preserving
 diffeomorphisms and quantum groups (algebras). One
 can study the geometric action on the coadjoint orbit
 of $w_{\infty}$ or $W_{\infty}$
 as it was  discussed in \cite{anpzv}, \cite{smm}.
 These actions  are  relevant to both $w$ gravity and two-dimensional
 hydrodynamics (the finite dimensional analog of the trigonometric
 FFZ algebra in an ideal  two-dimensional hydrodynamics
  was considered in \cite{z}).
  It is unclear  if there  is a hidden
 quantum symmetry of these  geometrical actions.

 It is known that there is a
 $W_{\infty}$  in the $c=1$ strings
 and corresponding matrix models \cite{kpw}-\cite{matrix}.
 What quantum algebra (if any) can be constructed in
 this model and what is the natural value of a deformation
 parameter $q$ in this case ? Will this quantum symmetry
  exist in the case of deformed $c=1$ model, for example
 in the case of two-dimensional black hole ?

 It is also known that the action for
 pure Yang-Mills theory in
 two dimensions is invariant under the
  area-preserving
 diffeomorphisms (see, for example \cite{witten2qcd}).
 Can one find a quantum symmetry in this case ?

 Let us finally note that recently \w has been discussed in a
 context of a bosonization of current-current interactions
 \cite{dima}.  In the framework of this approach the
 quantum symmetry  may appear in some new condensed matter
 problems.

We hope to return to these and related questions
 in the following publications.

\medskip

\medskip

\noindent
{\bf Acknowledgments.}

\noindent
 I would like to thank A.Bilal, D.Haldane,  D. V. Khveshchenko, I.R.Klebanov,
  A.M. Polyakov  and
 P.B. Wiegmann
 for interesting  discussions.  This work was supported by the National
 Science Foundation grant NSF PHY90-21984.

\noindent
{\bf Note added}
\noindent

  After this paper has been  submitted for
 publication  I  was aware about recent
 preprint \cite{sato} where the quantum group symmetry in
  Landau problem and in quantum Hall system  had  been
 also discussed.
\newpage
\renewcommand{\thesection}{A}
\renewcommand{\thesubsection}{A.\arabic{subsection}}

\vspace{12mm}
\pagebreak[3]
\setcounter{section}{1}
\setcounter{equation}{0}
\setcounter{subsection}{0}
\setcounter{footnote}{0}

\newpage
{\renewcommand{\Large}{\normalsize}

\end{document}